\documentclass[reprint,preprintnumbers,superscriptaddress,amsmath,amssymb,prb]{revtex4-1}
\usepackage{braket}
\usepackage{miller}
\usepackage{changes}
\usepackage[textsize=tiny]{todonotes}
\usepackage{graphicx}
\usepackage[utf8]{inputenc}
\usepackage{multirow}
\usepackage{float}
\usepackage{bm}
\usepackage{verbatim}
\usepackage{xfrac}
\usepackage{array}
\newcolumntype{C}[1]{>{\centering\let\newline\\\arraybackslash\hspace{0pt}}m{#1}}
\newcolumntype{L}[1]{>{\raggedright\let\newline\\\arraybackslash\hspace{0pt}}m{#1}}
\usepackage{siunitx}

\begin{document}
	
	\title{Lasing in topological edge states of a 1D lattice}
	
	\author{P. St-Jean}
	\email{philippe.saintjean@c2n.upsaclay.fr}
	\affiliation{Centre de Nanosciences et de Nanotechnologies, CNRS, Univ. Paris-Sud, Université Paris-Saclay, C2N - Marcoussis, 91460 Marcoussis, France}

	\author{V. Goblot}
	\affiliation{Centre de Nanosciences et de Nanotechnologies, CNRS, Univ. Paris-Sud, Université Paris-Saclay, C2N - Marcoussis, 91460 Marcoussis, France}
	
	\author{E. Galopin}
	\affiliation{Centre de Nanosciences et de Nanotechnologies, CNRS, Univ. Paris-Sud, Université Paris-Saclay, C2N - Marcoussis, 91460 Marcoussis, France}
	
	\author{A. Lemaître}
	\affiliation{Centre de Nanosciences et de Nanotechnologies, CNRS, Univ. Paris-Sud, Université Paris-Saclay, C2N - Marcoussis, 91460 Marcoussis, France}
	
	\author{T. Ozawa}
	\affiliation{INO-CNR BEC Center and Dipartimento di Fisica, Università di Trento, I-38123 Povo, Italy}
	
	\author{L. Le Gratiet}
	\affiliation{Centre de Nanosciences et de Nanotechnologies, CNRS, Univ. Paris-Sud, Université Paris-Saclay, C2N - Marcoussis, 91460 Marcoussis, France}
	
	\author{I. Sagnes}
	\affiliation{Centre de Nanosciences et de Nanotechnologies, CNRS, Univ. Paris-Sud, Université Paris-Saclay, C2N - Marcoussis, 91460 Marcoussis, France}
	
	\author{J. Bloch}
	\affiliation{Centre de Nanosciences et de Nanotechnologies, CNRS, Univ. Paris-Sud, Université Paris-Saclay, C2N - Marcoussis, 91460 Marcoussis, France}
	
	\author{A.~Amo}
	\affiliation{Centre de Nanosciences et de Nanotechnologies, CNRS, Univ. Paris-Sud, Université Paris-Saclay, C2N - Marcoussis, 91460 Marcoussis, France}
	
	\date{\today}
	%
	%
	%
	\maketitle
	
	
	
	\onecolumngrid
	\noindent\textbf{Topology describes properties that remain unaffected by smooth distortions. Its main hallmark is the emergence of edge states localized at the boundary between regions characterized by distinct topological invariants. This feature offers new opportunities for robust trapping of light in nano- and micro-meter scale systems subject to fabrication imperfections and to environmentally induced deformations. Here we show lasing in such topological edge states of a one-dimensional lattice of polariton micropillars that implements an orbital version of the Su-Schrieffer-Heeger Hamiltonian. We further demonstrate that lasing in these states persists under local deformations of the lattice. These results open the way to the implementation of chiral lasers in systems with broken time-reversal symmetry and, when combined with polariton interactions, to the study of nonlinear topological photonics.}\\ \\
	\twocolumngrid
	
	\noindent Topological phase transitions in condensed matter have been extensively studied over the last decade. A key manifestation of these transitions is the emergence, at the frontier between materials exhibiting distinct topological phases, of localized states that are unaffected by disorder. One example of this topological protection is provided by chiral edge states at the surface of topological insulators that allow unidirectional transport immune to backscattering\cite{Hasan2010}.
	
	Initially proposed by Haldane and Raghu\cite{Haldane2008}, the idea of extending these topological arguments to the realm of photonics has recently triggered considerable efforts to engineer optical devices that are unaffected by local perturbations and fabrication defects\cite{Lu2014}. For example, topological properties have been used to create polarization-dependent unidirectional waveguides \cite{Sollner2015a}, optical delay lines with enhanced transport properties\cite{Mittal2014}, backscattering-immune chiral edge states \cite{Wang2009, Hafezi2011, Hafezi2013, Rechtsman2013}, and protected bound states in parity-time-symmetric crystals\cite{Poli2015, Weimann2017}.
	
	The emergence of edge states at the boundary between materials with distinct topological invariants provides an efficient way to create localized photonic modes whose existence is protected by topology
	\cite{Rechtsman2013}. Lasing in these kind of modes would then be robust against fabrication defects, local deformations caused by temperature or other unstable ambient conditions, and long term degradation, all of which would eventually result in the modification of the local optical potential~\cite{Pilozzi2016}. The main difficulty that has prevented the observation of lasing in topological modes is the need to implement topological lattices in media exhibiting optical gain. In this sense, microcavity polaritons, mixed quasiparticules formed from the strong coupling between cavity photons and quantum well excitons\cite{Carusotto2013}, provide a unique platform: they allow for low-threshold lasing\cite{Deng2002, Kasprzak2006}, even at room temperature\cite{Christopoulos2007, Kena-Cohen2010}, and the engineering of topological properties in lattices of resonators\cite{Milicevic2017,Baboux2016a}. 
	
	In this work we report lasing in topological edge states of a 1D lattice of coupled semiconductor micropillars. The lattice implements an orbital version of the Su-Schrieffer-Heeger (SSH) model by coupling $l\ne 0$ polariton modes confined in a zigzag chain of micropillars. Under non-resonant optical pumping, we show that gain occurs in the topological states localized at the edges of the chain. Then, taking advantage of polariton nonlinearities, we demonstrate the topological robustness of the lasing action against optically induced lattice deformations. These results open the way to the realisation of edge lasers of arbitrary geometry, in which the lasing mode would be determined by the boundary between topologically distinct regions, regardless of its shape.
	\\

	\begin{figure*}
		\centering
		\includegraphics[trim=0cm 0cm 0cm 0cm, width=\textwidth]{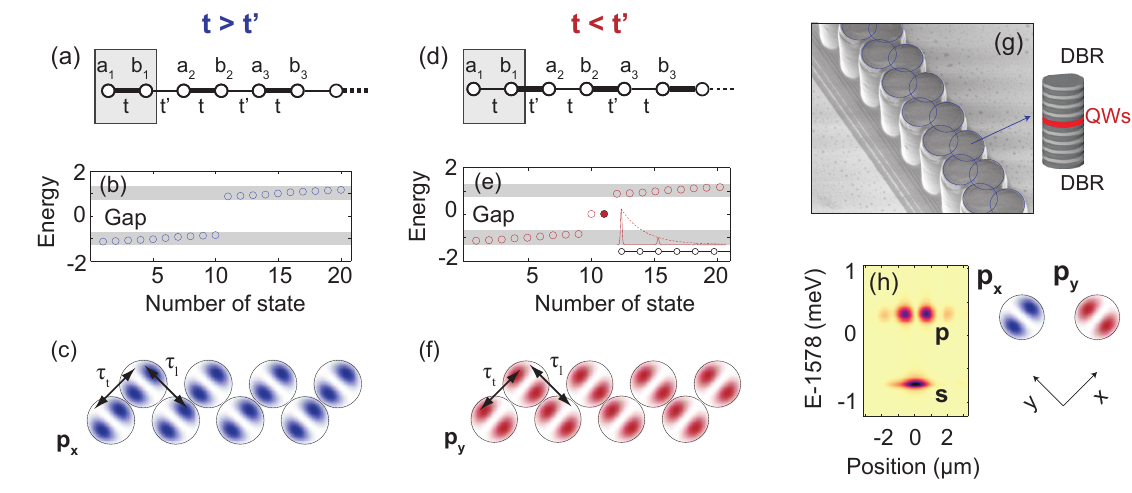}
		\caption{(a) and (d) Schematic representation of the two dimerizations in the SSH model. The unit cell is indicated by the gray rectangles. (b) and (e) Band structure associated to both dimerizations; the gray areas represent the spectral position of the bands. The inset in Panel (e) shows the distribution of the wave-function (solid line) and of the envelope function (dashed line) over the first six sites of the lattice for the gap state indicated by the filled circle. (c) and (f) Schematic representations of the $p_{x}$ and the $p_{y}$ sub-spaces. Each sup-space exhibits a distinct dimerization determined by the orientation of the longitudinal ($t_{l}$) and transversal ($t_{t}$) couplings at the edge of the chain. (g) Scanning electron microscopy (SEM) image of a zigzag chain of coupled micropillars etched out of a cavity; the blue circles are added for visibility. The enlargement shows a schematic representation of a single pillar embedding quantum wells (QWs) between distributed Bragg reflectors (DBRs). (h) Spectrally resolved real space emission of an isolated micropillar showing $s$- and $p$-orbitals for polaritons (schematic representation of the $p_{x,y}$ modes is presented on the right side).}
		\label{tightBinding}
	\end{figure*}

	\noindent\textbf{The orbital SSH model}
	
	\noindent To engineer topological edge states in a 1D lattice, we implement an orbital version of the SSH model. The standard SSH model describes a one-dimensional lattice with two sites per unit cell and different intra- ($t$) and inter-cell ($t'$) hopping amplitudes. Within the tight-binding approximation, such a model is captured by a Hamiltonian with chiral symmetry (see section 2 of Supplementary information):
	
	\begin{equation}
		\label{Hamiltonian}
		\hat{H}=\sum_{j} t\hat{a}_{j}\hat{b}^{\dagger}_{j} + t'\hat{a}^{\dagger}_{j}\hat{b}_{j-1} + h.c.,
	\end{equation}
	
	\noindent where $\hat{a}^{\dagger}_{j}$ ($\hat{b}^{\dagger}_{j}$) are the creation operators on the sub-lattice site $a_{j}$ ($b_{j}$) in the $j^{th}$ unit cell. Its eigenstates form two bands ($\pm$) in momentum space separated by a gap of magnitude $2\left|t-t'\right|$. The eigenfunctions in the $\left\{a, b\right\}$ sub-lattice basis take the form\cite{Delplace2011} $\left|u_{k,\pm}\right\rangle = \left(1/\sqrt{2}\right)\left(e^{-i\phi\left(k\right)}, \pm 1\right)^{\dagger}$.
	
	$\hat{H}$ exhibits two topologically distinct phases associated to the two possible dimerizations $t>t'$ and $t<t'$. The different topology of these two phases is revealed by considering the winding $\mathcal{W}$ of the phase $\phi\left(k\right)$ across the Brillouin zone:
	
	\begin{equation}
		\label{Zak}
		\mathcal{W} = \frac{1}{2\pi} \int_{BZ} {\frac{\partial \phi\left(k \right) }{\partial k}} d k
	\end{equation}
	
	
	\noindent which corresponds to the Zak phase divided by $\pi$. 
	
	Although the value of $\mathcal{W}$ associated to either dimerization depends on the definition of the unit cell, in finite size chains the choice is unambiguous since $t$ is defined by the hopping amplitude between the first and second sites of the chain. Under this definition, the $t>t'$ or $t<t'$ dimerizations exhibit respectively strong and weak coupling between the edge pillars and the rest of the chain (as depicted in Fig. \ref{tightBinding} (a) and (d)), and correspond to the trivial ($\mathcal{W} = 0$) and non-trivial ($\mathcal{W} = 1$) topological phases~\cite{Delplace2011}. Band structures calculated for chains of 20 sites exhibiting $\mathcal{W} = 0$ and $\mathcal{W} = 1$ are presented in Fig. \ref{tightBinding} (b) and (e), respectively. The most notable difference is the existence in the latter case of two states localized at the center of the energy gap corresponding to topological states localized at each end of the chain. The distribution of the wave-function over the first six pillars for the eigenstate indicated by the filled circle, is presented in the inset of Panel (e). Its envelope (dashed line) decays as $\lambda^{n}$, where $n$ is the unit cell number counted from the edge and $\lambda=t/t'$.
	
	To implement the SSH Hamiltonian, we consider the collective photon modes of a 1-dimensional lattice of coupled polariton micropillars. The photonic modes of a single micropillar are confined in the three dimensions of space leading to discrete energy levels: the ground state ($s$) exhibits a cylindrical symmetry along the growth axis, and the first excited states ($p_{x,y}$) present two degenerate antisymmetric orbitals orthogonal to each other (see Fig.~\ref{tightBinding} (h)). The orbital version of the SSH model considered in this work relies on the coupling of these $p$-orbitals in a one dimensional lattice of micropillars arranged in a zigzag configuration (Fig.~\ref{tightBinding}~(g)).
	
	In zigzag chains, $p_{x}$ and $p_{y}$ orbitals are respectively oriented along the diagonal and anti-diagonal axes (see Fig. \ref{tightBinding} (c) and (f)), and the hopping amplitude between consecutive micropillars strongly depends on the orientation of the axis linking these pillars\cite{Jacqmin2014}. The coupling is typically an order of magnitude stronger for orbitals oriented along the hopping direction than for orbitals oriented perpendicular to the hopping\cite{Jacqmin2014}. We define these different hopping strengths as longitudinal ($t_{l}$) and transverse ($t_{t}$), respectively. Note that $p_{x}$ and $p_{y}$ form independent sub-spaces, as the symmetry of the orbitals prevents the coupling between adjacent orthogonal orbitals. Consequently, if we consider the sub-space of $p_{x}$ modes (Fig.~\ref{tightBinding}~(c)), photons are subject to an alternate hopping strength as we move along the chain. This corresponds exactly to the SSH model described in Hamiltonian~(\ref{Hamiltonian}), where the intra- and inter-cell hopping strength are respectively $t_{l}$ and $t_{t}$. The large hopping anisotropy ($t_{l}>>t_{t}$) leads to the opening of a significant energy gap, as depicted in Fig. \ref{tightBinding}~(b). This sub-space exhibits strong coupling ($t_{l}$) between the first two (and last two) pillars, and does not present edge states; on the other hand, the $p_{y}$ sub-space (Fig.\ref{tightBinding}~(e)-(f)) starts and ends with the weak coupling $t_{t}$ and contains a topological edge state at each end of the chain. Similar two-fold sub-spaces have been studied using the polarization-dependent hopping of $s$ modes in zigzag chains of photonic resonators\cite{Solnyshkov2016, Kruk2017}.\\

	\begin{figure}
		\includegraphics[trim=0cm 0cm 0cm 0cm, clip, width=83mm]{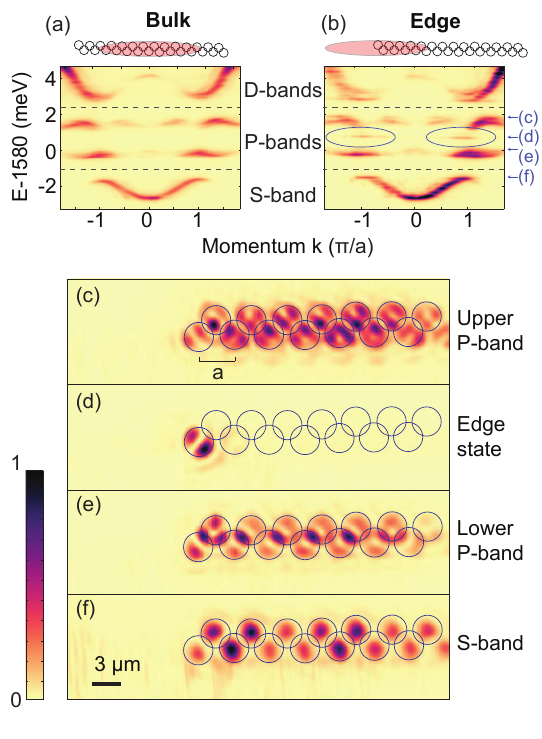}
		\caption{(a) and (b) Momentum space image of the emission from the bulk (a) and from the edge (b) of a zigzag chain ($\mathrm{a=3.4~\mu m}$ is the lattice periodicity, as depicted in Panel (c)). The area of the chain which is detected and excited is schematically represented on top of each figure. (c)-(f) Spatial images of the PL intensity at the energies of the upper (c) and lower (e) P-bands, of the S-band (f), and of the orbital gap state (d). The excitation spot used for the spatial images is located over the edge of the chain, as depicted on top of panel (b). The positions of the micropillars are indicated by the blue circles; blue arrows in Panel (b) indicate the spectral position at which each image is taken.}
		\label{spectre}
	\end{figure}
	
	\noindent\textbf{Imaging orbital states in the linear regime}
	
	\noindent To implement this orbital SSH model we etched zigzag chains of coupled micropillars from a planar microcavity containing 12 GaAs quantum wells sandwiched between two Ga$_{0.05}$Al$_{0.95}$As$/$Ga$_{0.8}$Al$_{0.2}$As Bragg mirrors with 32 (top) and 40 (bottom) pairs (see Methods and Ref.~[\onlinecite{Sala2015}] for further details). We can probe the orbital states by exciting the system nonresonantly with a continuous wave single mode laser focused on an elliptical spot of $\mathrm{2~\mu m}$ width and $\mathrm{50~\mu m}$ length (FWHM) that covers 25 unit cells of the lattice (see Fig. \ref{spectre} (a)). The system presents a splitting between linear polarizations oriented along and perpendicular to the main axis of the chain~\cite{Sturm2014}. For each of these two polarizations the system presents identical characteristics. We select the emission linearly polarized parallel to the long axis of the chain (similar results were obtained in the orthogonal polarization).

	Figure~\ref{spectre}(a) shows the photoluminescence intensity as a function of emission energy and momentum (directly proportional to the angle of emission) for low power excitation ($\mathrm{200~\mu W}$). It reveals three families of energy bands. The lowest band (S-band) arises from the coupling of $s$-mode polaritons. Due to their cylindrical symmetry, the hopping amplitude between these modes is completely isotropic ($t=t'$), and no energy gap is observed. At higher energies, we observe two sets of bands formed from the coupling of $p$-modes. They are separated by an energy gap of 1.7 meV due to the orbital hopping anisotropy $t_{l}>>t_{t}$. 
	At even higher energies, additional bands arising from the coupling of $d$-modes can be seen.
	
	When moving the elliptical spot over the edge of the chain (see Fig. \ref{spectre} (b)), we observe the emergence of a discrete state in the middle of the P-bands energy gap (indicated by blue circles), as expected for the topological edge states associated to the $p_{y}$ sub-space. Figure~\ref{spectre} (d) displays the real space emission at the energy of the edge mode, attesting its strong localization at the end of the chain. 
	In contrast, the S- and P-bands exhibit emission over the whole excited area (Fig.~\ref{spectre}~(c),(e),(f)). 
	
	The gap state wave-function exhibits three features that unambiguously demonstrate its topological nature. Firstly, we can see from the geometry of this edge state that it is formed from orbitals belonging to the topologically non-trivial $p_{y}$ sub-space. 
	Secondly, we observe that this state is strongly localized on the last pillar: the intensity in the second unit cell is an order of magnitude weaker than in the first. This can be understood from the low ratio $t_{t}/t_{l}$, which sets the penetration depth of the topological edge state. From the measured size of the gap and the amplitude of the bands, we estimate $t_{t} / t_{l}$ to be 0.15, consistent with the observed localization in the outermost pillar. Finally, we observe at the energies of the upper and lower P-bands (Fig. \ref{spectre} (c) and (e)) that the wave-function in the edge pillar only exhibits orbital modes belonging to the trivial sub-space $p_{x}$. This is consistent with tight-binding calculations that show that the bulk states vanish at the edge pillars in the topologically non-trivial $p_{y}$ sub-space (see section 1 of Supplementary information).\\
	
	\begin{figure}
		\includegraphics[trim=0cm 0cm 0cm 0cm, clip, width=83mm]{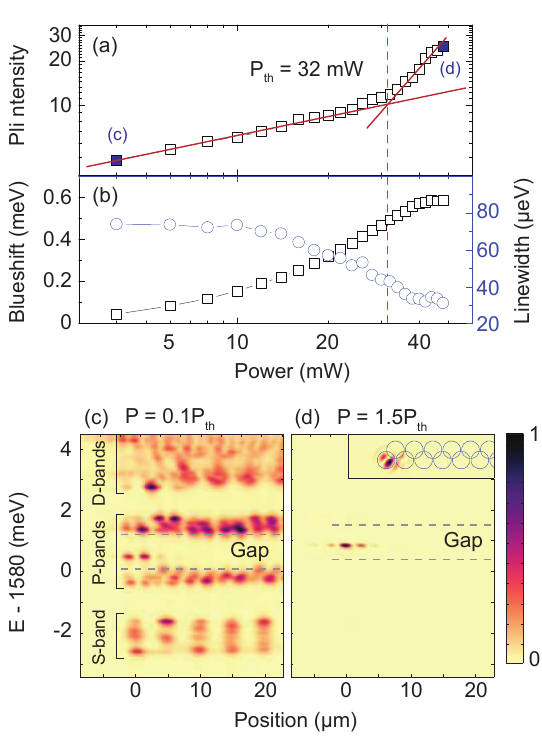}
		\caption{(a) Integrated intensity, and (b) linewidth (black squares) and blueshift (blue circles) of the emission from the topological edge state as a function of the excitation power. The lasing threshold is indicated by the dashed red line. Excitation is provided by an elongated sport centered over the edge of the chain, similar to that depicted in Fig. \ref{spectre} (b). (c)-(d) Normalized PL intensity as a function of energy and longitudinal position along the chain for an excitation power below (c) and above (d) the lasing threshold, corresponding to data points indicated by blue squares in Panel (a). The orbital band gap is indicated by horizontal dashed lines. The inset shows a spatial image of the PL at the energy of the edge state (the position of the micropillars is indicated by blue circles).}
		\label{longSpotLasing}
	\end{figure}
	
	\noindent\textbf{Lasing in the topological edge state}
	
	\noindent One interesting feature of polariton micropillars in the strong-coupling regime is the ability to trigger lasing in excited states, such as the topological edge states described above. This is possible thanks to the driven-dissipative nature of cavity polaritons: the steady-state lasing is determined by the interplay of pumping intensity, nonlinear polariton relaxation and emission lifetime\cite{Richard2005, Wouters2008, Baboux2016}. Both the relaxation rate and the lifetime are strongly influenced by the photon-exciton detuning~\cite{Levrat2010}, that is, the energy difference between the bare photon and exciton modes that couple to form polaritons.
	

	
	To achieve polariton lasing in the edge states of our orbital SSH chain, we select an exciton-photon detuning of -9.4~meV (see Methods), which favours relaxation of polaritons in the $p$-band states. Figure~\ref{longSpotLasing} (a), presents the spatially-integrated PL intensity at the energy of the edge state as a function of the excitation power. A non-linear increase of the intensity is observed at a threshold power of $\mathrm{P_{th}=32~mW}$, indicating the triggering of lasing in the topological edge state. Simultaneously, the linewidth of the emission collapses evidencing the increase of temporal coherence characteristic of the lasing regime (blue circles in Fig.~\ref{longSpotLasing}~(b)). Energy resolved real-space images (Fig.~\ref{longSpotLasing}~(d)) show that for $\mathrm{P=1.5 P_{th}}$ the emission from the edge state completely overcomes that of the bulk bands. The inset shows that the localization of the edge state is well preserved in the lasing regime. The observed energy blueshift in Fig.~\ref{longSpotLasing}~(d) with respect to (c) (and reported in (b) for all excitation powers) arises from the presence of a reservoir of excitons injected by the excitation laser that rigidly shifts the whole band structure under the excitation spot (dashed lines show the position of the central gap of the P-bands).
	
	The fact that the chain lases preferentially at the edge state rather than at the bulk P-bands states can be explained by the localized character of this mode. Indeed, polaritons in band states can propagate away from the excitation spot, reducing their lifetime and precluding lasing in these modes in favor of the confined edge mode.\\
	
	
	
	\begin{figure*}
		\includegraphics[trim=0cm 0cm 0cm 0cm, clip, width=\textwidth]{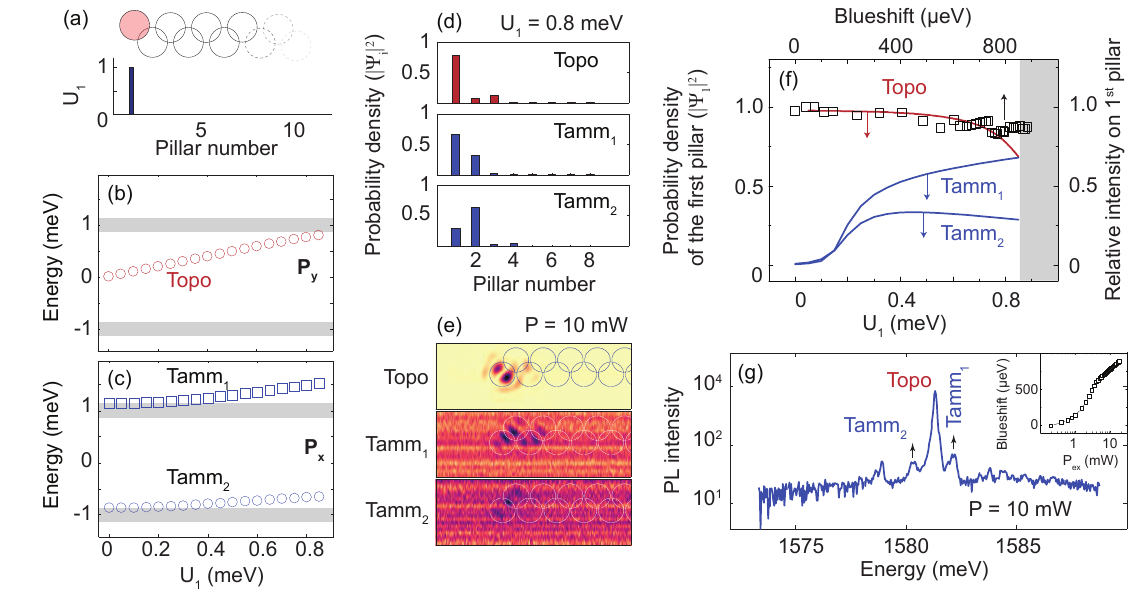}
		\caption{(a) Schematic representation of the distribution of on-site energies considered for the theoretical investigation of a localized perturbation. (b)-(c) Calculated energies of topological ($\mathrm{Topo}$ in the $p_{x}$ sub-space) and non-topological ($\mathrm{Tamm_{1,2}}$ in $p_{y}$) edge states of the locally perturbed SSH Hamiltonian, as a function of the on-site energy of the first pillar $U_{1}$. Gray areas indicate the positions of the upper and lower P-bands. (d) Squared amplitude of the calculated wave-functions for the three edge states presented in Panels (b) (Topo) and (c) ($\mathrm{Tamm_{1,2}}$) for $U_{1}=0.8~meV$. (e) Spatial distribution of the PL at the energies of Topo, $\mathrm{Tamm_{1}}$ and $\mathrm{Tamm_{2}}$ for an excitation localized over the edge pillar, with a power of $\mathrm{P=3P_{th}=10~mW}$. The colorscale is identical to that used in Fig. \ref{spectre}. (f) Calculated evolution of the localization (defined as the wave-function squared amplitude at the first pillar) of the topological ($\mathrm{Topo}$, red line) and non-topological ($\mathrm{Tamm_{1,2}}$, blue lines) edge states as a function of the perturbation energy $U_{1}$. Black squares present the measured relative intensities of the PL from Topo at the position of the first pillar as a function of its spectral blueshift; the position of the upper P-band is indicated by the gray area. (g) PL spectrum for the same excitation conditions as in Panel (e). The inset shows the evolution of the measured blueshift of the topological edge state as a function of the excitation power.}
		\label{lasingSmall}
	\end{figure*}

	\noindent\textbf{Robustness of topological lasing}
	
	We now investigate the robustness of the topological lasing mode against local deformations of the lattice. The SSH Hamiltonian (Eq. \ref{Hamiltonian}) presents a chiral symmetry, that is, it anticommutes with the $\sigma_{z}$ Pauli matrix. The main consequence of this is that the topological mode appears in the middle of the gap. Since this symmetry is preserved when considering disorder in the hopping strengths, the energy and localization of the topological mode are immune to this type of disorder (see section 2 of Supplementary information). Therefore, the kind of local perturbations to which this mode is most sensitive to are changes of on-site energies, which break the chiral symmetry, especially in the first lattice site since the strong localization of the wave-function (Fig.~\ref{spectre}~(d)) mitigates the effect of energy perturbations in other sites.
	
	To evaluate theoretically the effect of such energy perturbation in the first pillar, we add an on-site energy term $U_{1}a_{1}^{\dag} a_{1}$ in the Hamiltonian presented in Eq. \ref{Hamiltonian} (Fig. \ref{lasingSmall} (a)). By diagonalizing this perturbed Hamiltonian for a chain of 20 pillars, we can evaluate the evolution of the eigenergy and eigenfunction of the topological edge mode in the $p_{y}$ sub-space as a function of $U_{1}$ (we used $t_{l}\mathrm{=1~meV}$ and $t_{t}\mathrm{=0.15~meV}$, which reproduce the experimentally observed P-bands and gap). The main effect of the perturbation $U_{1}$ is to modify the energy of the edge mode in the gap, as depicted in Fig.~\ref{lasingSmall}~(b). Remarkably, its spatial localization is hardly affected: even for a perturbation energy $U_{1}=0.8~\mathrm{meV}$, where the edge state is almost resonant with the upper P-band, the wave-function is strongly localized over the first pillar (see Fig.~\ref{lasingSmall}~(d)). This is further evidenced by the red line in Fig. \ref{lasingSmall} (f) showing a high probability density of the edge state on the first pillar when varying $U_{1}$ from 0 up to 0.85~meV, at which point the edge state merges with the bulk band. 
	
	To experimentally test the robustness of lasing in the edge state against perturbations of the energy of the edge pillar, we use the nonlinear properties of polaritons. Under non-resonant excitation, a highly populated exciton reservoir is formed at an energy 11 meV above the lasing mode. This reservoir interacts repulsively with low energy polaritons inducing their blueshift with a magnitude controlled by the excitation power~\cite{Bajoni2008}. The inset of Figure~\ref{lasingSmall}~(g) shows the energy of the edge state emission as a function of the excitation power, when using a small pump spot of $\mathrm{3.5~\mu m}$ (FWHM) in diameter localized on the first pillar. The injected exciton reservoir continuously increases the local energy of the first pillar, resulting in the blueshift of the lasing mode. In agreement with the simulations, the localization length is immune to the perturbation as we observe that the emission remains highly confined in the first pillar all over the whole power dependence (see Fig.~\ref{lasingSmall}~(e), squares in Fig. \ref{lasingSmall}~(f), and section 3 of Supplementary information for the detailed lasing characteristics of the localized excitation scheme). This behavior is a clear evidence of the robustness of the topological mode, in the $p_{y}$ sub-space, to local perturbations. 
	
	In contrast, the topologically trivial $p_{x}$ sub-space presents a very different behavior. Simulations of the perturbed Hamiltonian for the trivial dimerization show that the energy shift of the first pillar results in the emergence of two localized modes (labeled $\mathrm{Tamm_{1,2}}$ in Fig. \ref{lasingSmall} (c)) with energies in the central gap and above the upper P-band, respectively. They correspond to non-topological Tamm modes which have been extensively discussed in the context of surface states induced by symmetry-breaking perturbations\cite{Zak1985, Malkova2009, Blanco-Redondo2016}. Contrary to the topological edge state, their wave-function is extended over a few pillars (Fig. \ref{lasingSmall} (d)) and their distribution strongly varies with the value of $U_{1}$. This behavior is visible in Fig. \ref{lasingSmall} (f) (blue lines), showing a low calculated weight of their wavefunction on the edge micropillar. Experimentally, 
	they appear in the zigzag chain as a weak emission (side peaks on both side of the topological state in Fig.~\ref{lasingSmall}~(g)). Their spatial distribution only contains $p_{x}$ components and extends significantly more than the topological edge mode, as seen in Fig.~\ref{lasingSmall} (e).
	
	Note that despite the fact that the localized reservoir induces the emergence of Tamm modes in the $p_{x}$ sub-space of the zigzag chain, lasing takes place dominantly in the topological $p_{y}$ edge mode over the explored power range. This is a consequence of the stronger localization of the topological mode, which overlaps more efficiently with the small pump spot than the Tamm modes, and thus presents a lower lasing threshold. A fundamental difference between the topological and Tamm modes in the SSH chain is that the former appears as a single state in the gap, while the latter appear in pairs for $U_{1} \neq 0$. In a lattice containing only the trivial dimerization, the Tamm states would then be prone to multimode effects~\cite{Blanco-Redondo2016}.\\
	
	\noindent\textbf{Conclusion}
	
	The observation of lasing in a topologically protected edge state presented here provides a direct demonstration of robust light trapping in topological structures. Our results open the way to lasing in modes with more complex geometries. By engineering the boundary between media described by distinct topological invariants we envision edge lasers with arbitrary shape\cite{Harari2016}. A particularly interesting case is that of lasing in chiral edge modes in systems with broken time-reversal symmetry, recently predicted in the context of polaritons\cite{Nalitov2015, Karzig2015}. In addition, the nonlinear control of the optical landscape employed in the present work demonstrates the suitability of cavity polaritons for exploring nonlinear phenomena in topological photonics\cite{Hadad2016}.\\

	\noindent\textbf{Methods}\\
	\noindent\textit{Sample description}\\
	\noindent The zigzag chain of coupled micropillars is etched out of a planar semiconductor cavity consisting of a $\mathrm{Ga_{0.05}Al_{0.95}As}$ $\lambda /2$ layer embedded between two $\mathrm{Ga_{0.05}Al_{0.95}As/Ga_{0.2}Al_{0.8}As}$ Bragg mirrors formed from 28 (40) pairs in the top (bottom) mirror. Twelve GaAs quantum wells of 7~nm width are grown at the central maxima of the electromagnetic field in the cavity, resulting in strong photon-exciton coupling exhibiting a 15~meV Rabi splitting. After the epitaxy, the cavity is processed by electron beam lithography and dry etching to form a zigzag chain of overlapping cylindrical micropillars: the diameter of the pillars (3~$\mu\mathrm{m}$) overcomes the center-to-center distance (2.4~$\mu\mathrm{m}$), allowing for the hopping of polaritons\cite{Galbiati2012}. The distance between consecutive pillars is constant and the orientation of the axis linking pillars alternates between + and - $\mathrm{45^{o}}$ with respect to the length of the chain; these orientations are defined by the axes $\hat{x}$ and $\hat{y}$, respectively (see Fig. \ref{tightBinding} (b)).
	
	An exciton photon detuning of $\mathrm{\delta=-9.4~meV}$ is chosen in order to favor lasing in P-bands. This detuning is defined as $\delta = E_{c}(0) - E_{X}(0)$, where $E_{c}(k)$ and $E_{X}(k)$ describe respectively the energy dispersion of S-mode cavity photons and quantum well excitons as a function of their in-plane momentum $k$.
	\\
	
	\noindent\textit{Experimental technique}\\
	\noindent Non-resonant PL measurements were realized with a single-mode CW laser at 754~nm. The elongated spot is engineered using a cylindrical lens. The emission is collected through a microscope objective and imaged on the entrance slit of a spectrometer coupled to a CCD camera with a spectral resolution of $\mathrm{\sim30~\mu eV}$. Real- and momentum-space PL images are realized by imaging the sample surface and the Fourier plane of the objective, respectively. A $\lambda / 2$ wave-plate and a polarizer are used to select emission polarized either along or across the long axis of the chains. The sample is cooled down at $\mathrm{T=4~K}$.\\
	
	\noindent\textbf{Acknowledgements}
	
	We thank G. Montambaux for fruitful discussions. This work was supported by the French National Research Agency (ANR) project ”Quantum Fluids of Light” (ANR-16-CE30-0021) and program Labex NanoSaclay via the project ICQOQS (Grant No. ANR-10-LABX-0035), the French RENATECH network, the ERC grant Honeypol and the EU-FET Proactiv grant AQUS (Project No. 640800). P.S.-J. acknowledges the financial support of the Natural Sciences and Engineering Research Council of Canada (NSERC).\\

	\newpage
	\onecolumngrid
	\noindent\textbf{\LARGE Supplementary information}

	\section{Tight-binding calculations of the orbital wave-functions}
	Eigenstates of the $p_{x,y}$ sub-spaces can be theoretically calculated by diagonalizing the Hamiltonian presented in Eq. 1 of the main text. In the case at hand, $t$ and $t'$ are replaced by $\mathrm{t_{l}=1~meV}$ and $\mathrm{t_{t}=0.15~meV}$ for the $p_{x}$ sub-space, and vice-versa for the $p_{y}$ sub-space (these values are extracted from the direct measurement of the flatness and energy gap of the P-bands). For a chain formed from $N=20$ micropillars, each sub-space is formed from 20 eigenstates distributed in two conduction bands separated by a band gap of 1.7~meV (see Fig.~1~(b) and (e) of the main text).\\
	
	As presented extensively in the main paper, two gap states in the $p_{y}$ sub-space are localized at each end of the chain as a consequence of the non-trivial topology associated to the dimerization $t'>t$. The localization of these gap states can be visualized by considering the distribution of their wave-function over every site of the chain, as presented in the top and middle panel of Fig. \ref{tightBinding} (a).\\
	
	Another important consequence of the non-trivial topology is that the wave-functions of the 18 bulk states (i.e. states with energies in the conduction bands) strngly vanishes at the edge pillars. This is clearly seen by summing the wave-functions squared amplitude of the 18 bulk states. The distribution of this summation over every pillar of the chain is presented in the bottom pannel of Fig. \ref{tightBinding} (a), and clearly shows a vanishing value at the position of the edge pillars.\\
	
	\begin{figure*}[h]
		\centering
		\includegraphics[trim=0cm 0cm 0cm 0cm, width=\textwidth]{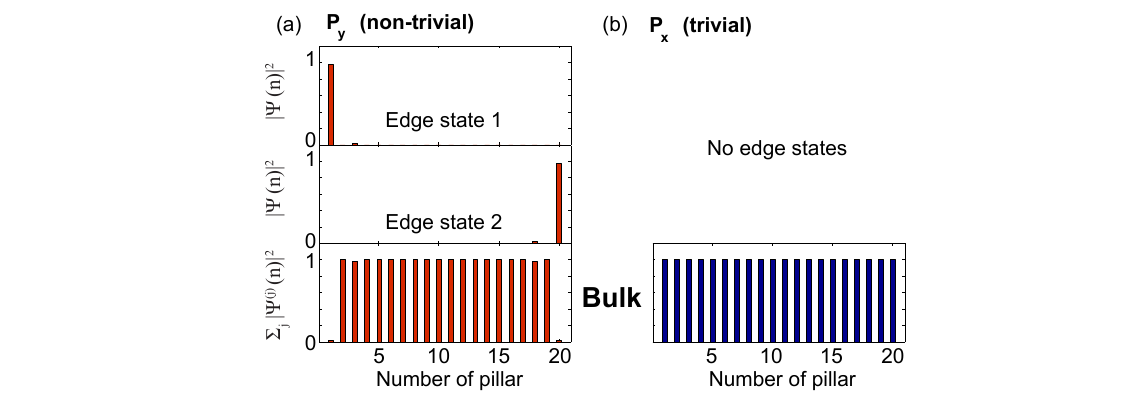}
		\caption{(a) Top and middle panel: distribution over every pillar of the wave-function squared amplitude of both gap states. Bottom panel: summation of the wave-function squared amplitudes of all 18 bulk states of the $p_{x}$ sub-space. (b) Similar for the $p_{y}$ sub-space, where there are no edge states.}
		\label{tightBinding}
	\end{figure*}
	
	For the $p_{x}$ sub-space, the edge pillars are strongly coupled to the rest of the chain resulting in the topologically trivial phase of the SSH model. Therefore, there are no gap states in this sub-space, and the 20 eigenstates are delocalized over every pillar including the edge pillars (see Fig. \ref{tightBinding} (b)).\\
	
	\section{Chiral symmetry of the SSH Hamiltonian}
	
	A system exhibits chiral symmetry if there exists a unitary transformation $\mathcal{U}_{c}$ that anticommutes with the Hamiltonian\cite{Chiu2016}: $\{H,\mathcal{U}_{c}\}=0$. For an infinite SSH lattice, the Hamiltonian of the system (defined in Eq. 1 of the main text) can be expressed in momentum space by a $2\times2$ matrix
	
	\begin{equation}
	H(k)=\vec{d}(k)\cdot\vec{\sigma},
	\label{hamiltonian}
	\end{equation}
	
	where $\sigma_{x,y,z}$ are the Pauli matrices, and 
	
	\begin{eqnarray}
	d_{x}(k)=t+t'cos(ka)\nonumber\\
	d_{y}(k)=t'sin(ka)\\
	d_{z}(k)=0,\nonumber
	\end{eqnarray}
	
	\noindent with $a$ the unit cell dimension\cite{Delplace2011}.\\
	
	The SSH Hamiltonian thus possesses a chiral symmetry defined by the Pauli matrix $\sigma_{z}$: $\{H(k),\sigma_{z}\}=0$. The principal consequence of this symmetry is that the energy spectrum is symmetric around the origin: every eigenstate $\ket{u_{E}}$ with eigenenergy $E$ has a partner $\ket{u_{-E}}=\sigma_{z}\ket{u_{E}}$ with energy $-E$. For states with energy $E=0$ (such as the topological edge states considered in this work), these states are their own partner ($E=-E=0$), and their spectral position is not affected by perturbations, as long as $\{H(k),\sigma_{z}\}=0$.\\
	
	For example, perturbations that affect the hopping strengths $t$ and $t'$ only modify off-diagonal terms in Hamiltonian \ref{hamiltonian} and thus do not affect chiral symmetry. Consequently, the spectral positions of zero-energy states (such as the topological edge states of the $p_{y}$ sub-space discussed in this work) are immune to such perturbations. This can be seen by diagonalizing a Hamiltonian similar to that used for calculating the band structure presented in Fig. 1 (e) of the main text, but where we include random fluctuations to the hopping strengths: $t_{l,t}^{(i)} \rightarrow t_{l,t}^{(i)}(1+\alpha r_{l,t}^{(i)})$, where $r_{l,t}^{(i)}$ are random numbers between -1 and 1, $\alpha$ describes the amplitude of the hopping fluctuations, and $i$ refers to the site number along the chain.\\
	
\begin{figure*}[h]
		\centering
		\includegraphics[trim=0cm 0cm 0cm 0cm, width=\textwidth]{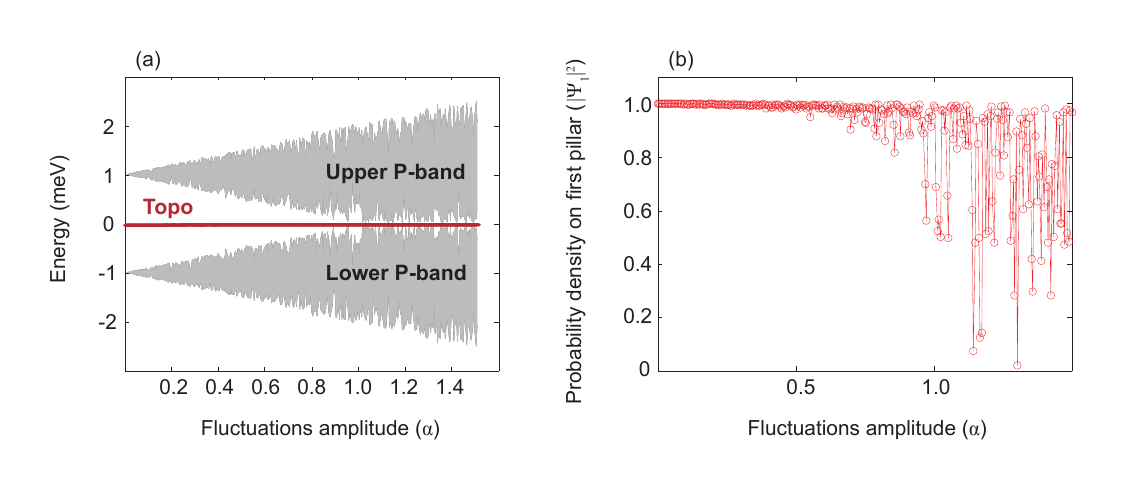}
		\caption{Evolution of the energy (a) and probability density on first pillar (b) of the $p_{y}$ topological edge state as a function of the amplitude of the hopping fluctuations. Gray areas in panel (a) indicate the width of the P-bands.}
		\label{fluctuations}
	\end{figure*}

	Fig. \ref{fluctuations} (a) indeed shows that the spectral positions of the $p_{y}$ topological edge state (red line) is not affected by fluctuating hopping strengths. However, the width of the upper and lower P-bands, indicated by gray areas, increases constantly following a stochastic evolution. This increase is due to the statistical loss of contrast between $t_{t}$ and $t_{l}$, which also leads to an increase of the penetration depth of the topological edge state (which is related to the ratio $t_{t}/t_{l}$) as presented in Fig. \ref{fluctuations} (b). The localization of the edge state however never completely vanishes under such perturbations.\\
	
	Thus, perturbing hopping strengths does not provide the most compelling platform for investigating the robustness of the topological edge modes, because chiral symmetry of the SSH Hamiltonian insures that their energy will not be affected. On the other hand, perturbations that break chiral symmetry by adding diagonal terms to the Hamiltonian, such as considering second-neighbor hopping or different on-site energies, have more drastic consequences as they modify the spectral position of the topological edge state. This is the case in our work, where we locally break chiral symmetry by modifying the on-site energy of the edge pillar. The consequences of this perturbation are presented in the section \textit{Robustness of topological lasing} of the main text.\\
	
	\section{Polariton lasing using a localized excitation}
	
\begin{figure*}[h]
		\centering
		\includegraphics[trim=0cm 0cm 0cm 0cm, width=\textwidth]{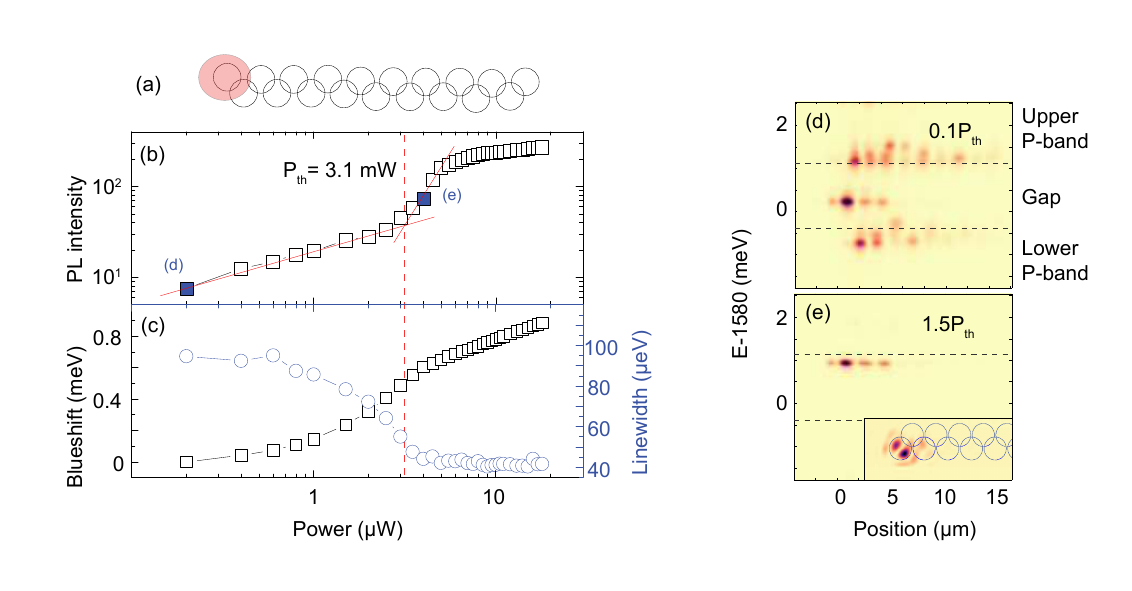}
		\caption{(a) Schematic representation of the localized excitation scheme. (b) Integrated PL intensity as a function of the excitation power. The red dashed line indicates the position of the lasing threshold. (c) Measured PL linewidth (blue circles) and spectral blueshift (black squares) of the topological edge state as a function of the excitation power. (d)-(e) PL intensity as a function of emission energy and position along the chain for an excitation power below (d) and above (e) the lasing threshold; the excitation power for each measurement is indicated by the blue squares in Panel (b). Inset in Panel (e) shows a spatial image of the PL at the energy of the topological edge state.}
		\label{lasing}
	\end{figure*}

	Under a localized excitation (the situation shown in Fig. 4 of the main text), the lasing threshold is reduced by an order of magnitude with respect to the case of an extented excitation spot (Fig. 3 of the main text), down to $\mathrm{P_{th}=3.1~mW}$. Above this threshold, a non-linear increase of the emission is observed in the I-P curve presented in Fig. \ref{lasing} (b). PL images as a function of position and emission energy still show that, in the lasing regime, emission from the topological edge state completely overcomes that from bulk bands (see Fig. \ref{lasing} (d)-(e)). As well, the topological edge state is almost completely localized over the last pillar as depicted in the inset, even at the strongest excitation power used ($\mathrm{P_{ex}=6P_{th}=18~mW}$). Similarly as with the large excitation spot, polariton lasing leads to a narrowing of the emission linewidth; on the other hand, interactions with the localized reservoir are more important, leading to a stronger spectral blueshift (both effects are respectively presented by the blue circles and black squares in Fig. \ref{lasing} (c)).\\
	
	For the large excitation spot, we observed a rigid shift of all the band structure: of bulk bands as well as of the edge state. Under a localized excitation spot, the chiral symmetry is broken as described in the previous section, and the induced blueshift of the last pillar only changes the spectral position of the edge state. Thus the topological mode sweeps across the energy gap as the excitation power is increased, as shown in Fig. \ref{lasing} (d) and (e)).\\
	
\end{document}